\documentclass[11pt,twoside]{article}
\usepackage{graphicx,floatflt,amssymb,epsfig,epsf}
\usepackage{times}
\usepackage{a4wide}
\usepackage{cite}

\usepackage{fancyheadings}
\advance \headheight by 3.0truept       
\newlength{\nseparation}
\newenvironment{nfigure}[1]
        {\begin{figure}[#1]\hrule\vspace{\nseparation}\par}
        {\vspace{\nseparation}\par \hrule \end{figure}}

\def\bra {\langle}
\def\ket {\rangle}
\def\be {\begin{equation}}
\def\ee {\end{equation}}
\def\bea {\begin{eqnarray}}
\def\eea {\end{eqnarray}}

\def\issue(#1,#2,#3){{\bf #1}, #2 (#3)}
\def\opcit(#1){ {\em op. cit.}, #1}
\def\etal {\em et al.}

\def\APP(#1,#2,#3){Acta Phys.\ Polon.\ \issue(#1,#2,#3)}
\def\ARNPS(#1,#2,#3){Ann.\ Rev.\ Nucl.\ Part.\ Sci.\ \issue(#1,#2,#3)}
\def\CPC(#1,#2,#3){Comp.\ Phys.\ Comm.\ \issue(#1,#2,#3)}
\def\CIP(#1,#2,#3){Comput.\ Phys.\ \issue(#1,#2,#3)}
\def\EPJC(#1,#2,#3){Eur.\ Phys.\ J.\ C\ \issue(#1,#2,#3)}
\def\EPJD(#1,#2,#3){Eur.\ Phys.\ J. Direct\ C\ \issue(#1,#2,#3)}
\def\IEEETNS(#1,#2,#3){IEEE Trans.\ Nucl.\ Sci.\ \issue(#1,#2,#3)}
\def\IJMP(#1,#2,#3){Int.\ J.\ Mod.\ Phys. \issue(#1,#2,#3)}
\def\JHEP(#1,#2,#3){J.\ High Energy Physics \issue(#1,#2,#3)}
\def\JPG(#1,#2,#3){J.\ Phys.\ G \issue(#1,#2,#3)}
\def\MPL(#1,#2,#3){Mod.\ Phys.\ Lett.\ \issue(#1,#2,#3)}
\def\NP(#1,#2,#3){Nucl.\ Phys.\ \issue(#1,#2,#3)}
\def\NIM(#1,#2,#3){Nucl.\ Instrum.\ Meth.\ \issue(#1,#2,#3)}
\def\PL(#1,#2,#3){Phys.\ Lett.\ \issue(#1,#2,#3)}
\def\PRD(#1,#2,#3){Phys.\ Rev.\ D \issue(#1,#2,#3)}
\def\PRL(#1,#2,#3){Phys.\ Rev.\ Lett.\ \issue(#1,#2,#3)}
\def\SJNP(#1,#2,#3){Sov.\ J. Nucl.\ Phys.\ \issue(#1,#2,#3)}
\def\ZPC(#1,#2,#3){Zeit.\ Phys.\ C \issue(#1,#2,#3)}
\def\EPL(#1,#2,#3){Europhys.\ Lett.\ \issue(#1,#2,#3)}

\newcommand{\bbs}{\ensuremath{B_s\!-\!\Bbar{}_s\,}}
\newcommand{\bbms}{\bbs\ mixing}
\newcommand{\bbd}{\ensuremath{B_d\!-\!\Bbar{}_d\,}}
\newcommand{\bbmd}{\bbd\ mixing}
\newcommand{\dg}{\ensuremath{\Delta \Gamma}}
\newcommand{\dm}{\ensuremath{\Delta M}}
\newcommand{\BsorBsbar}{\raisebox{7.7pt}{$\scriptscriptstyle(\hspace*{8.5pt})$}
  \hspace*{-10.7pt}\!\Bbar_{s}}
\newcommand{\eq}[1]{Eq.~(\ref{#1})}
\newcommand{\eqsand}[2]{Eqs.~(\ref{#1}) and (\ref{#2})}

\newcommand{\sgn}{\mbox{sign}\,}
\newcommand{\Bbar}{\,\overline{\!B}}

\newcommand{\nn}{\nonumber \\}
\newcommand{\ov}[1]{\overline{#1}}
\newcommand{\lt}{\left}
\newcommand{\rt}{\right}

\begin{document}
\thispagestyle{plain}
TTP07-37
\hfill 
arXiv:0801.0143 [hep-ph]\\
CU-PHYSICS/15-2007
\hfill 
December 2007

\begin{center}
\boldmath
{\Large \bf Resolving the sign ambiguity in $\dg_s$ with 
$B_s \to D_s K$}\\
\unboldmath
\vspace*{1cm}
\renewcommand{\thefootnote}{\fnsymbol{footnote}}
Soumitra Nandi${}^{1,2}$ and Ulrich Nierste${}^1$ \\
\vspace{10pt}
{\small
    ${}^{1)}$ {\em Institut f\"ur Theoretische Teilchenphysik --
               Karlsruhe Institute of Technology, 
               Universit\"at Karlsruhe, D-76128 Karlsruhe, Germany} \\
    ${}^{2)}$ {\em Department of Physics, University of Calcutta, 92 A.P.C.
        Road, Kolkata 700009, India}}
 
\normalsize
\end{center}


\begin{abstract}
The analysis of tagged $B_s\to J/\psi \phi$ decays determines 
the CP phase $\phi_s$ in \bbms\ with a two--fold ambiguity. The solutions
differ in the sign of $\cos \phi_s$ which equals the sign of the width
difference $\dg_s$ among the two $B_s$ mass eigenstates. We point out 
that this ambiguity can be removed with the help of $B_s \to D_s K$ 
decays. We compare untagged and tagged strategies and find the tagged
analysis more promising. The removal of the sign ambiguity in $\dg_s$ 
can be done with relatively low statistics and could therefore
be a target for the early stage of $B_s \to D_s K$ studies. 
\end{abstract}

The theoretical description of \bbms\ involves the elements $M^s_{12}$ and
$\Gamma^s_{12}$ of the mass and decay matrices, respectively \cite{run2}. The
precise measurement of the mass difference $\dm_s=M_H-M_L$ between the heavy
and light mass eigenstates of the $B_s$ system determines $|M^s_{12}|$
\cite{teva}.  Yet theoretical uncertainties still permit new--physics
contributions to $|M^s_{12}|$ of order 30\%. The new contributions to
$M_{12}^s$ can even exceed the Standard Model value in magnitude, because
Standard--Model and new--physics contributions generally come with an
arbitrary relative complex phase. To probe new physics in $M^s_{12}$ further,
it is therefore mandatory to study the phase of $M^s_{12}$ experimentally.
The width difference between the heavy and light mass
eigenstates of the $B_s$ system is given by%
\be%
\dg_s = \Gamma_L - \Gamma_H = 2 |\Gamma^s_{12}| \cos\phi_s \qquad
\mbox{with }\quad \phi_s = \arg
\left(- \frac{M^s_{12}}{\Gamma^s_{12}}\right) . \label{defphi}%
\ee%
Since $\Gamma^s_{12}$ is unaffected by new physics, measurements of
$\phi_s$ probe new physics in $M^s_{12}$. It is useful to decompose
$\phi_s$ as $\phi_s=\phi_s^{\rm SM}+\phi_s^\Delta$ with the two terms
denoting the Standard--Model and new--physics contributions to $\phi_s$,
respectively.  Currently most experimental information on
$\phi_s^\Delta$ stems from the decay $B_s \to J/\psi \phi$. The CP phase
in this decay is the difference $\phi_s^\Delta -2 \beta_s$ between $\arg
M^s_{12}$ and twice the phase of the $b\to c\ov c s$ decay amplitude
with $\beta_s= \arg [- V_{ts} V_{tb}^*/(V_{cs} V_{cb}^*)]\; =\; 0.020
\pm 0.005 \; = \; 1.1^\circ \pm 0.3^\circ $. While it is safe to neglect
$\phi_s^{\rm SM}=(4.2\pm 1.4 )\cdot 10^{-3} \; = \; 0.24^\circ \pm
0.08^\circ$ \cite{bbln,ln} and to identify $\phi_s$ with
$\phi_s^\Delta$, we keep $2 \beta_s$ non--zero in our formulae.  The
untagged decay $\BsorBsbar \to J/\psi \phi$ provides information on
$\dg_s \cos(\phi_s^\Delta-2\beta_s)$ and $|\sin
(\phi_s^\Delta-2\beta_s)|$ \cite{dgexp}. These measurements have been
combined with experimental constraints on the semileptonic CP asymmetry
$a_{\rm sl}$ \cite{asl} to determine the allowed ranges for $\phi_s^\Delta$
\cite{ln,combined}. Recently, the CDF collaboration has presented a
tagged analysis of $B_s \to J/\psi \phi$ \cite{cdftagged} with the two
solutions
\be
\phi_s^\Delta-2\beta_s \,\in \, \lt[ -1.36,-0.24  \rt] 
\qquad\mbox{or }\qquad
\phi_s^\Delta-2\beta_s \, \in \,  \lt[ -2.90,-1.78  \rt] \qquad 
 @68\% \, \mbox{CL} .
\label{cdfphi}
\ee
The quoted ranges correspond to the analysis in Ref.~\cite{cdftagged}
which constrains $\dg_s$ in \eq{defphi} with the theoretical value
$|\Gamma_{12}|=0.048\pm 0.018\, \mbox{ps}^{-1} $ \cite{bbln,ln,bbgln}.
\eq{defphi} implies $\sgn \dg_s = \sgn \cos\phi_s$, so that the two
solutions in \eq{cdfphi} correspond to $\dg_s>0$ and $\dg_s<0$,
respectively.  Lifetime measurements in the components of the angular
distributions of an untagged $B_s \to J/\psi \phi$ sample determine
$|\dg_s|$, which implies a four--fold ambiguity in $\phi_s^\Delta$
\cite{yuval,uli1,ln}.  Neglecting the small $\beta_s$ for a moment, all
quantities which can be extracted from $B_s \to J/\psi \phi$ and also
$a_{\rm sl}$ suffer from the same two--fold ambiguity $\phi_s^\Delta
\leftrightarrow \pi - \phi_s^\Delta$ visible in \eq{cdfphi}.  Thus at
present we do not have any information on the sign of $\dg_s$. The two
solutions in \eq{cdfphi} correspond to different values of $\cos
\delta_{1,2}$, where $\delta_1$ and $\delta_2$ are strong phases. In
order to resolve the ambiguity in $\sgn\cos\phi_s=\sgn\dg_s$ one must
determine $\sgn\cos \delta_{1,2}$.  If this is done with naive
factorisation \cite{naive}, the second solution in \eq{cdfphi} is
obtained. However, if the strong phases measured in $B_d \to J/\psi K^*$
\cite{babarphases} are used, one finds the first solution in \eq{cdfphi}
(see the discussion in \cite{cdftagged}). It should be noted that the
$SU(3)_{\rm F}$ symmetry links $B_d \to J/\psi K^*$ only partially to
$B_s \to J/\psi \phi$. Only the component of the $\phi$ meson with
U--spin equal to 1 belongs to the symmetry multiplet of the $K^*$.  The
decay amplitude into the equally large U--spin--zero component cannot be
related to $B_d\to J/\psi K^*$.
Since there is also no reason to trust naive factorisation in $B_s \to J/\psi
\phi$, we conclude that the sign ambiguity in $\dg_s$ is unresolved.
 
The tagged decays $\BsorBsbar \to D_s^\mp K^\pm $ were proposed to
determine the angle $\gamma$ of the unitarity triangle (UT). They do not
involve any penguin pollution and are therefore hadronically very clean
\cite{adk}.  However, these decays are also sensitive to a possible new
phase in \bbms\ and really determine $\phi_s^\Delta-2 \beta_s + \gamma$
(up to a tiny correction of order $0.1^\circ$). An exhaustive study of
$\BsorBsbar \to D_s^\mp K^\pm $ including the effects of a non--zero
$\phi_s^\Delta$ can be found in Ref.~\cite{fleischer}, which also
focuses on the determination of $\gamma$ assuming that $\phi_s^\Delta$
has been determined unambigously with other methods. In this paper we
propose to view $\BsorBsbar \to D_s^\mp K^\pm $ from a different angle:
We exploit that $\gamma$ is well--measured at B factories and show that
$\BsorBsbar \to D_s^\mp K^\pm $ can be used to discriminate between the
two solutions with $\cos \phi_s >0 $ and $\cos \phi_s <0 $ in
\eq{cdfphi}. This information can be found with relatively low
statistics, much before studies of this decay mode at LHCb become
competitive for the determination of $\gamma$.  The resolution of the
sign ambiguity in $\dg_s$ can be achieved either with a lifetime
measurement in untagged $\BsorBsbar \to D_s^\mp K^\pm $ decays or by
inspecting the sign of the oscillating term in a tagged $B_s \to D_s^\mp
K^\pm $ data sample.  We will discuss both strategies below.  Since the
CDF experiment has already gathered more than 100 $B_s \to D_s^\mp K^\pm
$ events \cite{cdfdsk}, the determination of $\sgn \dg_s$ maybe even
within reach of the Tevatron.

$\gamma$ is well--known from the decay $B_d \to \rho^+\rho^-$, which
measures the UT angle $\alpha$ plus a potential phase of new physics in
\bbmd: If the measured CP asymmetries in $B_d \to \rho^+\rho^-$ and
$B_d\to J/\psi K_s$ are combined to solve for $\gamma=\pi-\alpha-\beta$,
any new physics in \bbmd\ drops out. Exploiting the smallness of the
penguin pollution in $B_d \to \rho^+\rho^-$ and using QCD factorisation
\cite{bbns} Ref.~\cite{beneke} finds $\gamma = 71^{\circ} \pm
5^{\circ}$. For simplicity we define
\be%
\gamma_s \;\equiv\; \gamma \, -\,  2\beta_s \;= \; 69^{\circ} \pm 5^{\circ}.   
\label{defgs}
\ee%
Alternatively one can include other $B \to \rho\rho$ modes and use
isospin symmetry to control the penguin pollution \cite{brrpeng}. In
principle this analysis comes with discrete ambiguities for $\gamma$ as
well. However, the global analysis of the UT only permits the one
solution for $\gamma$ quoted above.

The time--dependent $\BsorBsbar \to D_s^\mp K^\pm $ decay rates 
involve \cite{run2,adk,fleischer}
\bea%
\lambda_{D^-_s K^+} &=&  {q\over p} 
    \frac{\bra D^-_s K^+ | \Bbar_s \ket}{\bra D^-_s K^+ | B_s \ket }
 \,=\, \big| \lambda_{D^-_s K^+} \big| \, 
        e^{-i \, (\gamma_s + \phi_s^\Delta- \delta) } \nn 
\lambda_{D^+_s K^-} &=& {q\over p} 
    \frac{\bra D^+_s K^- | \Bbar_s \ket}{\bra D^+_s K^- | B_s \ket }   
 \,=\, \frac{1}{\big| \lambda_{D^-_s K^+} \big|} \, 
        e^{-i \, (\gamma_s + \phi_s^\Delta + \delta) } . \nonumber 
\eea%
Here $q/p$ encodes \bbms\ in the usual way and $\delta$ is a strong
phase which equals zero if the matrix elements are computed in the
factorisation approximation \cite{fleischer}.  The $\BsorBsbar \to
D^{\mp}_s K^{\pm}$ decays are colour--allowed tree level processes and
the factorisation approximation is exact in the limit of a large number
$N_c$ of colours. We point out that the only $1/N_c$ corrections to the
matrix elements stem from annihilation topologies which are empirically
known to be small. The remaining corrections to the large--$N_c$ limit
are quadratic in $1/N_c$, see e.g.\ \cite{manohar}. Of course the
full-flesh tagged analysis can determine $\delta$ \cite{adk,fleischer},
but for our purposes it is sufficient to know that $\delta$ is small. We
conservatively assume $|\delta|<0.2$.

We introduce the shorthand notation 
\be%
b \, = \, \frac{2\, |\lambda_{D^-_s K^+}|}{1+ |\lambda_{D^-_s K^+}|^2}
 \label{defb}
\ee%
and note that $0< b \leq 1$. For realistic values 
$|\lambda_{D^-_s K^+}|\approx 0.4$ one finds $b\approx 0.7$. 

The time--dependent decay rates for the four relevant processes are 
\cite{run2}
\bea%
\Gamma\left(B_s(t) \to D^{\mp}_s K^{\pm}\right) \! &=& \! 
N e^{-\Gamma_s t} \Big[ \cosh({\Delta\Gamma_s \, t \over 2})  \pm    
     \lt( 1 - b\, \big| \lambda_{D^-_s K^+} \big| \rt) 
    \cos(\dm_s \, t) \nonumber \\
&& \hspace{-6ex}
-\; b \, \cos(\gamma_s + \phi_s^\Delta \mp \delta) 
            \sinh( {\Delta \Gamma_s \, t \over 2})  
 \; + \;    b\, 
    \sin(\gamma_s +\phi_s^\Delta  \mp  \delta )  
    \sin( \dm_s \, t ) \Big] ,\nn
\Gamma\left(\Bbar_s(t)  \to D^{\mp}_s K^{\pm}\right) \! &=&\! 
N e^{-\Gamma_s t} \Big[ \cosh({\Delta\Gamma_s \, t \over 2})  \mp    
     \lt( 1 - b\, \big| \lambda_{D^-_s K^+} \big| \rt) 
    \cos(\dm_s \, t) \nonumber \\
&& \hspace{-6ex}
-\; b \, \cos(\gamma_s + \phi_s^\Delta \mp \delta) 
            \sinh( {\Delta \Gamma_s \, t \over 2})  
 \; - \;    b\, 
    \sin(\gamma_s +\phi_s^\Delta  \mp  \delta )  
    \sin( \dm_s \, t ) \Big] .\quad 
\label{decayrt}
\eea%
Here $\Gamma_s=(\Gamma_L+\Gamma_H)/2$ and $N$ is a normalisation
constant.   
Now the untagged decay rate for the decay mode $\BsorBsbar \to 
D^{\pm}_s K^{\mp}$ reads
\bea%
\Gamma\left[D^{\mp}_s K^{\pm},t\right] &\equiv&  
\Gamma\left( B_s(t) \to D^{\mp}_s K^{\pm}\right) \; +\; 
\Gamma\left( \Bbar_s(t) \to D^{\mp}_s K^{\pm}\right) \nn
&=& 2 N e^{-\Gamma_s t}
\left[\cosh({\Delta\Gamma_s t \over 2}) \; -\; b\, 
  \cos(\gamma_s + \phi_s^\Delta \mp \delta)
  \sinh({\Delta\Gamma_s t \over 2}) \right],
\label{urate2}
\eea%
which is just the familiar two--exponential formula with the
time--dependent factors $\exp [-\Gamma_L t]$ and $\exp [-\Gamma_H t]$.
In practice one can determine two quantities from the untagged 
decay rate in \eq{urate2}, the branching fraction and the lifetime 
measured in the considered mode $\BsorBsbar \to D^{\pm}_s K^{\mp}$. 
The normalisation constant $N$ can be related to the CP--averaged 
branching fraction 
\cite{uli1,run2}:
\bea%
{\cal B} (\BsorBsbar \to D^{\mp}_s K^{\pm}) 
& \equiv &
{ {\cal B} (B_s \to D^{\mp}_s K^{\pm}) + 
  {\cal B} (\Bbar_s \to D^{\mp}_s K^{\pm}) \over 2} \nn
& =&
 {N \, \Gamma_s \over \Gamma_s^2 - (\dg_s)^2/4}
  \left[ 1  - b 
   \cos( \gamma_s + \phi^{\Delta}_s \mp \delta) 
   {\dg_s\over 2 \Gamma_s} \right] . 
\label{br} 
\eea%
From \eq{br} one finds
\bea%
{ {\cal B} (\BsorBsbar \to D^{+}_s K^{-}) 
- {\cal B} (\BsorBsbar \to D^{-}_s K^{+}) \over
  {\cal B} (\BsorBsbar \to D^{+}_s K^{-}) 
+ {\cal B} (\BsorBsbar \to D^{-}_s K^{+}) }
& =& b \,  
   { \sin( \gamma_s + \phi^{\Delta}_s) \, \sin \delta \over   
     1 - b\,\cos( \gamma_s + \phi^{\Delta}_s)  \cos \delta {\dg_s\over 2 \Gamma_s}} \, 
   {\dg_s\over 2 \Gamma_s} . \label{bras}
\eea%
Once this ratio of branching fractions is measured at the level 
of a few percent, it will be useful to place tighter bounds on 
$|\sin( \gamma_s + \phi^{\Delta}_s)  \sin \delta |$ and may help
the tagged analysis.  

For our purposes we need the lifetime information:  
A maximum likelihood fit of a time evolution given by \eq{urate2} to a
single exponential $\propto \exp [-\Gamma_{D_s^\mp K^\pm} t]$ determines
\cite{moser,uli1}: 
\be \Gamma_{D_s^\mp K^\pm} = \Gamma_s \; +\;  
      b\, \cos(\gamma_s + \phi_s^\Delta \mp \delta) \,   
  {\dg_s\over 2} \; = \;  \Gamma_s \; +\;
  b\, \cos(\gamma_s + \phi_s^\Delta \mp \delta) \cos  \phi_s^\Delta 
  \, \lt| \Gamma^s_{12} \rt|,
\label{life1}
\ee
where we neglected corrections of order
$(\dg_s)^2/\Gamma_s^2$. Comparing the rates $\Gamma_{D_s^+ K^-}$ and  
$\Gamma_{D_s^- K^+}$ gives the same information as \eq{bras}. More
important for us is the average of the two widths: Defining the quantity 
$L$ through  
\be%
{{\Gamma_{D^+_s K^-} + \Gamma_{{D^-_s K^+}}}\over 2} -  \Gamma_s \, =\,  
b \cos\delta \cos(\gamma_s +  \phi^{\Delta}_s) 
  \cos \phi^{\Delta}_s    \lt| \Gamma^s_{12} \rt|
\; \equiv \;  L\,  \lt| \Gamma^s_{12} \rt| ,
\label{lifeadd1}
\ee%
one first realises that the dependence of $L$ on $\delta$ is only
quadratic, so that the uncertainty from $\delta$ is inessential in view
of the error on $\gamma_s$ in \eq{defgs}: $|\delta|<0.2$ implies $0.98 <
\cos\delta <1$. Second we verify from \eq{lifeadd1} that we can resolve
the ambiguity in $\phi_s^\Delta$ by comparing the lifetime measured in
$\BsorBsbar \to D_s K$ with $1/\Gamma_s$, provided that $\phi_s^\Delta$
differs from $0$ or $\pi$.  This feature is illustrated in the left plot
of Fig.~\ref{fig}.
\begin{nfigure}{t}
\includegraphics[width=0.33\textwidth,angle=-90]{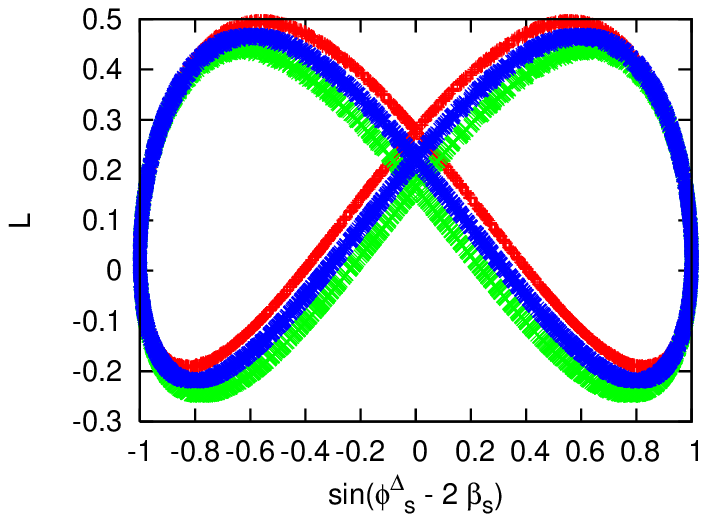}
~~~
\includegraphics[width=0.33\textwidth,angle=-90]{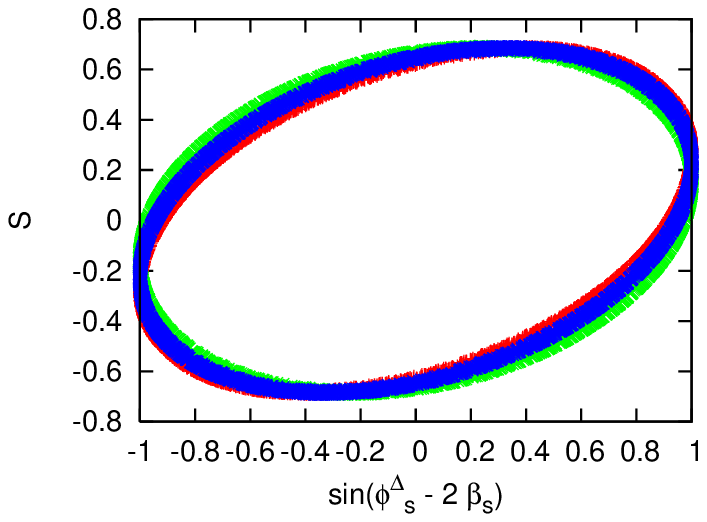}
\caption{The tagged 
  $\protect\BsorBsbar \protect\to \protect J/\psi\protect\phi$ analysis
  gives one solution for $\sin(\phi_s^\Delta-2\beta_s)$ corresponding to
  the two--fold ambiguity $\phi_s^\Delta -2\beta_s \leftrightarrow \pi-
  \phi_s^\Delta+2\beta_s $. The left plot shows how the measurement of
  $L$ in \eq{lifeadd1} can resolve this ambiguity. For $\sin
  (\phi_s^\Delta-2\beta_s)<0$ the upper branch corresponds to $\cos
  \phi_s^\Delta>0$ and $\dg_s>0$, while the lower branch corresponds to
  $\cos \phi_s^\Delta<0$ and $\dg_s<0$. For $\sin
  (\phi_s^\Delta-2\beta_s)>0$ the situation is reversed.
  The right plot shows the coefficient $S$ of the tagged analysis. The
  upper curve is for $\cos\phi_s>0$ meaning $\dg_s >0$, the lower curve 
  corresponds to $\cos\phi_s<0$ meaning $\dg_s <0$.     
  Both plots are for $b=0.7$ and $|\delta|<0.2$. The blue (dark) curves
  correspond to $\gamma=71^\circ$, the red (medium grey) and green
  (light) curves correspond to $\gamma=66^\circ$ and $\gamma=76^\circ$,
  respectively.\label{fig}}
\end{nfigure}
For example, the central values in the two intervals in \eq{cdfphi} both
correspond to $\sin(\phi_s^\Delta-2\beta_s)=-0.72$. But the solution
with $\dg_s>0$ comes with $L>0$, while $\dg_s<0$ implies $L<0$.  We do
not recommend to fit the data to a single exponential, because
\eq{life1} is only correct, if e.g.\ the experimental acceptance does
not vary with the decay length. Instead we propose to determine
$\phi_s^\Delta$ with the exact formula in \eq{urate2} with $\dg_s$
expressed as $\dg_s=2|\Gamma_{12}^s| \cos\phi_s^\Delta$ and
$|\Gamma_{12}^s|=0.048\pm 0.018\, \mbox{ps}^{-1}$ \cite{ln}.  Further
$\Gamma_s$ could be fixed to the theoretical value $\Gamma_s\simeq
1/\tau_{B_d}$.

Next we discuss the tagged analysis: With $\Gamma(B_s(t)\to D_s K) = 
[\Gamma(B_s(t)\to D_s^- K^+) + \Gamma(B_s(t)\to D_s^+ K^-)]/2$ we encounter
the CP asymmetry
\bea%
\!\!
\lefteqn{
{ \Gamma(B_s(t)\to D_s K) - \Gamma(\Bbar_s(t)\to D_s K)  \over
  \Gamma(B_s(t)\to D_s K) + \Gamma(\Bbar_s(t)\to D_s K) } \; =} 
  \nn
&& \qquad\qquad\qquad\qquad\qquad
{ b \cos\delta \sin(\gamma_s + \phi_s^\Delta) \sin(\dm_s t) \over
  \cosh (\dg_s t/2) \, -\,  
  b \cos\delta \cos(\gamma_s +  \phi^{\Delta}_s) \sinh (\dg_s t/2) } 
\quad \label{cpa} 
\eea%
The coefficient of the oscillating term,
\bea%
S &\equiv& b \cos\delta \sin(\gamma_s + \phi_s^\Delta)  \label{defs} 
\eea%
also permits the removal of the discrete ambiguity in $\phi_s^\Delta$,
because the replacement of $\phi_s^\Delta-2\beta_s$ by
$\pi-\phi_s^\Delta +2\beta_s$ changes $S$ dramatically, as shown in the
right plot of Fig.~\ref{fig}. $S$ even discriminates between the two
cases $\phi_s^\Delta=0$ and $\phi_s^\Delta=\pi$, which are the two
possible cases in the class of new physics models without new sources of
CP violation. 

Comparing the untagged and tagged method we find that discriminating the
two branches for $L$ in the left plot of Fig.~\ref{fig} means a lifetime
measurement with an accuracy of roughly $2\%$ requiring at least 2500
events, because the difference of the two solutions for $L
|\Gamma_{12}^s|/\Gamma_s$ hardly exceeds 0.04 and even vanishes if
$\phi_s^\Delta$ is close to 0 or $\pi$. The tagged measurement looks
better, even though tagging costs roughly a factor of 12--20 in
statistics. The two solutions with $\sin(\phi_s^\Delta-2\beta_s)=-0.72$
correspond to $S\simeq 0.3$ and $S\simeq -0.6$ and a fairly small data
sample should permit to discriminate between the two solutions. Finally
we remark that one can eliminate $\delta$ altogether, if both $L$ and
$S$ are measured precisely: \eqsand{lifeadd1}{defs} combine to
\be%
{\tan(\gamma_s + \phi^{\Delta}_s)} =  {S \over L } \,   
                       \cos{\phi^{\Delta}_s}
 \label{tan} .
\ee%
Now \eq{tan} has four solutions for $\phi^{\Delta}_s$ and two of them
can be eliminated with the information on the sign of $S$. The remaining
two solutions are not related by $\phi_s^\Delta -2\beta_s
\leftrightarrow \pi- \phi_s^\Delta+2\beta_s $, so that in combination
with $B_s \to J/\Psi \phi$ the discrete ambiguity is lifted with the
help of \eq{tan}.  We emphasize that \eq{tan} has been discussed before
in Ref.~\cite{fleischer}, where $\tan(\gamma_s + \phi^{\Delta}_s)$ is
expressed in terms of the coefficient of $\sinh(\dg_s t/2)$ in
\eq{decayrt}. The extraction of this coefficient has a sign ambiguity if
the sign of $\dg_s$ is unknown. Through \eqsand{lifeadd1}{tan} we have
merely expressed $\tan(\gamma_s + \phi^{\Delta}_s)$ in terms of $
|\Gamma^s_{12}|$ and $\cos{\phi^{\Delta}_s}$ to eliminate the implicit
dependence on $\sgn \dg_s$.

In conclusion we have discussed the removal of the two--fold ambiguity
in the extraction of $\phi_s^\Delta$ from tagged $B_s\to J/\psi \phi$
decays. We have shown that $B_s \to D_s^\pm K^\mp $ data  can be used 
to resolve this ambiguity. This analysis can be done with relatively 
low statistics, well before $B_s \to D_s^\pm K^\mp $ decays become
competitive for the determination of $\gamma$. Comparing untagged with
tagged analyses we find that the tagged analysis is more promising, 
despite of the penalty from small tagging efficiencies at hadron
colliders.  

We thank Gavril A.~Giurgiu and the referee for pointing out a misleading
typo in the first preprint version of this article.  SN~acknowledges the
hospitality of the TTP Karlsruhe and a grant from the Graduiertenkolleg
``Hochenergiephysik und Teilchenastrophysik''.  The work of UN is
supported by the BMBF grant 05 HT6VKB and by the EU Contract
No.~MRTN-CT-2006-035482, \lq\lq FLAVIAnet''.

\end{document}